# Magnetotransport Properties of Square-Net Compounds of NbSiSb and NbGeSb Single Crystals

Lei Guo[1,2,7], Weiyao Zhao[3,7], Ning Ding[1], Xin-Yao Shi[4], Meng Xu[2], Lei Chen[3], Guan-Yin Gao[5], Shuai Dong[1,8], and Ren-Kui Zheng[2,6,8]

[1] School of physics, Southeast University, Nanjing 211189, China

[2] State Key Laboratory of High Performance Ceramics and Superfine Microstructure, Shanghai Institute of Ceramics, Chinese Academy of Sciences, Shanghai 200050, China

[3] Institute for Superconducting & Electronic Materials, Innovation Campus, University of Wollongong, Wollongong, NSW 2500, Australia

[4] Suzhou Institute of Nano-Tech and Nano-Bionics, Chinese Academy of Sciences, Suzhou 215123, China

[5] Hefei National Laboratory for Physical Sciences at the Microscale, University of Science and Technology of China, Hefei 230026, China

[6] School of Materials Science and Engineering and Jiangxi Engineering Laboratory for Advanced Functional Thin Films, Nanchang University, Nanchang 330031, China

[7] These authors contribute equally to this work

[8] Author to whom any correspondence should be addressed.

E-mail: sdong@seu.edu.cn and zrk@ustc.edu

# Abstract

We successfully grew single crystals of Si- and Ge-square-net compounds of NbSiSb and NbGeSb whose excellent crystalline quality are verified using single-crystal x-ray diffraction θ-2θ scans, rocking curves, scanning and transmission electron microscopies. Since these two compounds share major crystallographic similarity with the topological nodal-line semimetals of ZrSiS family, we employ density functional theory (DFT) calculations and magnetotransport measurements to demonstrate their band structures as well as the electron scattering mechanisms. DFT calculations show that the fermiology shows strong anisotropy from the crystallographic *c*-axis to the *ab*-plane and weak anisotropy within the *ab* plane, which is consistent with the strong anisotropic magnetotransport behaviors. Following the Kohler's scaling rule we prove that similar interband and intraband electron-phonon scattering mechanisms work in both the NbSiSb and NbGeSb compounds. The study of electronic transport mechanism in the presence of external magnetic field renders deep insight into topological behavior together with it's Fermi surface, and the high similarity of crystallography and strong difference in band structures between the present single crystals and that of ZrSiS family provides the possibility to tune the band structure via element doping

(Some figures may appear in colour only in the online journal)

## 1. Introduction

Square nets of atoms, which can appear in a large variety of crystal structures, have been linked to many exciting physical properties, such as the superconductivity of cuprate- and iron-based superconductors and the band topology of topological semimetals [1-4]. Moreover, square-net compounds are relatively easy to be studied theoretically and appear in a great many of crystal structures, therefore, it becomes the basis for a variety of theoretical predictions of materials' properties. With the recent upsurge of research on topological quantum materials, many intermetallic compounds with square-net structures were demonstrated to be topological semimetals [5-7]. The most famous one is the topological nodal-line semimetal of ZrSiS which holds layered structure piled with ZrS layer and Si square-net layer and has attracted increasing attention for its long-range linear-dispersed band structure [8, 9]. For the ZrSiS compound the Si-square lattice's non-symmorphic symmetry protected nodal-line fermions dominates the quantum transport properties, resulting in ultrahigh mobility, butterfly shape magnetoresistance and Shubnikov-de Hass oscillations with nontrivial Berry phase. Similar quantum phenomena have been found in *MXY* (where *M* = Zr, Hf; *X* = Si, Ge, Sn; *Y* = S, Se, Te) [10-12] and isostructural LnSbTe (Ln = La, Ce) compounds, [13, 14] which are attributed to the *X*-square nets or Sb-square nets. Analogously, the Si- or Ge-square-net layers can be also found in the NbSiSb or NbGeSb compounds which share the same space group as that of the ZrSiS [5,15].

The distinguished Fermi surface and band structure leads to exotic magnetotransport phenomena in ZrSiS-family single crystals [8, 11, 16]. The magnetotransport has been employed to study the potential energy surfaces and the scattering mechanisms of semiconductors for several decades. Classically, the electrons in a conductor travel in straight lines between successive collisions under zero field. The magnetic field has significant effects on conductivity if it is strong enough to bend the trajectory appreciably during a free path, where the Lorentz force $e\mathbf{v}\times\mathbf{B}$ works. The electrons thus follow a circle orbit with the angular velocity of $\omega = eB/m^*$ ($m^*$ is the effective mass of an electron) under the magnetic field *B*, resulting in positive magnetoresistance. For a closed Fermi surface the electrons may convolute back to their initial positions if the magnetic field is strong enough, leading to the total *MR* to a saturation value. However, for an open Fermi surface, the electrons can keep going through the Brillouin zone boundary, resulting in increasing MR with no saturation [17]. In general, the study of electrical properties, especially large magnetoresistance effect ensures the further application of electronic devices in the future.

In this paper, we employ the chemical vapor transport (CVT) method to grow needle-like NbSiSb and NbGeSb single crystals and studied their band structures and magnetotransport properties. First-principles electronic structure calculations reveal the semimetal nature of the NbSiSb and NbGeSb crystals, which implies that both electron and hole pockets can be found in their Fermi surfaces. Large and non-saturating MR (over 200% at 3 K and 14 T) were observed for the NbSiSb, suggesting that the NbSiSb forms an open Fermi surface. For the NbGeSb the *MR* and Fermi surface follow similar trend, however, with a one-order-smaller *MR* value (~ 20%) at 3 K and 14 T. On the orders of magnitude, the magnetoresistance of these two materials can be comparable to the giant magnetoresistance (GMR) or colossal magnetoresistance (CMR) observed in metallic thin films and perovskite manganites [18].The study of the electric transport behavior of NbSiSb and NbGeSb will be a good supplement to the family of topological materials, which will help people understand the behavior of topological materials and it's Fermi surface more comprehensively. More recently, we noticed that the band topology related to the surface states of NbGeSb has been studied via angle resolved photoelectron spectroscopy, which suggests that this family is important in topological physics [19] and could be useful to instruct element-doping based band structure engineering among ZrSiS family.

## 2. Experimental Methods

High-quality NbSiSb and NbGeSb single crystals were grown by the CVT method using the iodine ($I_2$) as transport agent. Briefly, high-purity stoichiometric total amount (~1 g) of Nb,

Si or Ge and Sb powder (200 mesh), together with 20 mg/ml iodine, are sealed in quartz tube as the starting materials. The crystal growth was carried out in a two-zone furnace between 1050 °C (source) and 950 °C (sink) for 1 week. The as-grown single crystals [inserts of Fig 1(c,d)] exhibit needle-like shape with a typical size of approximately 2 to 6 mm in length along the *b*-axis direction and 1.5 mm$^2$ area in *ac* plane.

The crystal structure of the single crystals was characterized by a PANalytical X'pert x-ray diffractometer equipped with Cu $K\alpha_1$ radiation. High-resolution transmission electron microscopy (HRTEM) and selected area electron diffraction (SAED) were measured using a Tecnai G2F20 S-Twin transmission electron microscope. The chemical composition and distribution maps of the single crystals were measured via energy dispersive x-ray spectroscopy (EDS) using an x-ray energy dispersive spectrometer (Oxford Aztec X-Max80) installed on the Zeiss Supra 55 scanning electron microscope.

The electronic transport properties were measured by the standard four-probe method using a physical property measurement system (PPMS-14T, Quantum Design). Ohmic contacts were prepared on a fresh cleavage crysral surface using room-temperature cured silver paste. The electric current is parallel to the *b* axis while the direction of the magnetic field is parallel to the [101] direction of the crystal in the transverse magnetotransport measurement configuration.

The first-principles electronic structure calculations are performed using the Vienna *ab initio* simulation package (VASP) with the projector augmented-wave (PAW) potentials [20-22], based on the density functional theory (DFT) The exchange and correlation interactions were realized by the generalized gradient approximation (GGA), which is parametrized by Perdew, Burke, and Ernzerhof (PBE). In this calculation, energy cutoff was set to be 450 eV for the plane-wave basis and Brillouin zone (BZ) integration was sampled with a *k*-grid density of 7×7×3 using the Monkhorst-Pack *k*-points scheme. Besides, the lattice parameters and the ionic positions were fully optimized using a conjugate (CG) algorithm until the residual force on each atom was less than 0.01 eV/ Å where the convergence criteria for energy was set to be $10^{-5}$ eV.

## 3. Crystal Structure

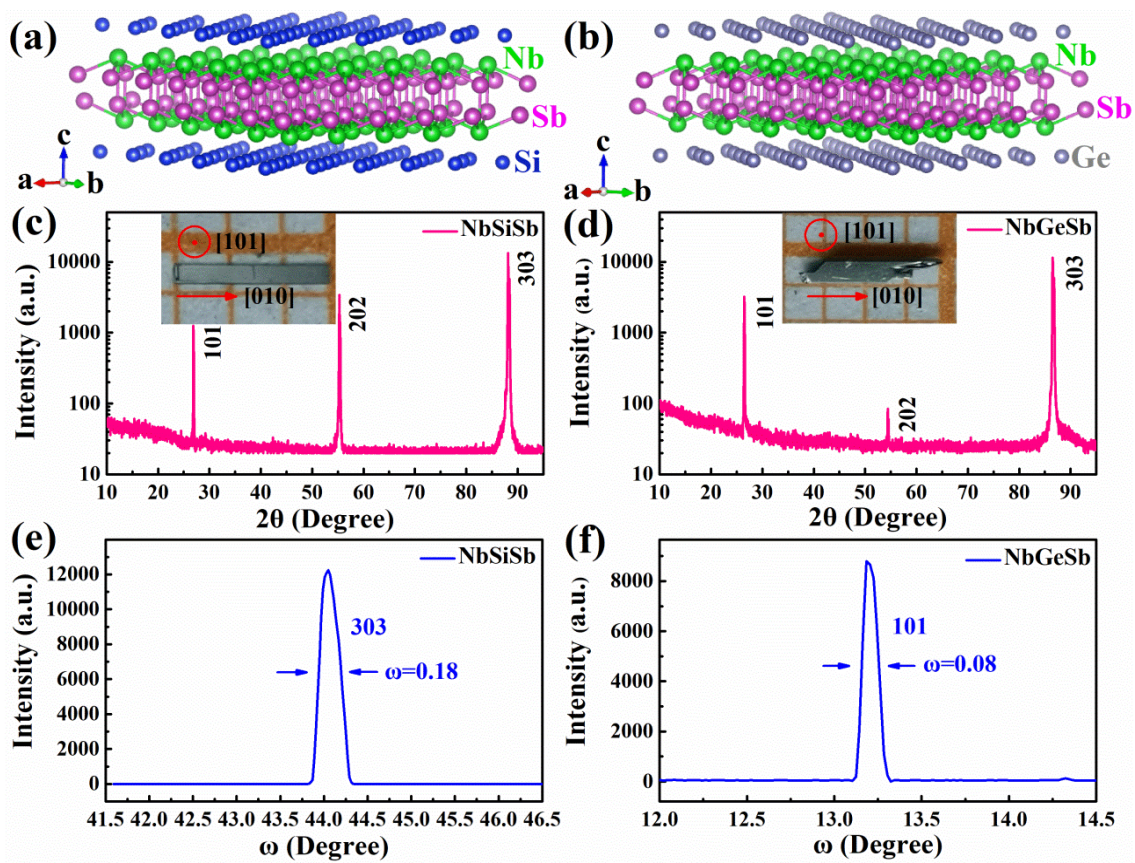

**Fig. 1** Crystallography properties of the NbSiSb and NbGeSb single crystals. (a,b) crystal structures. (c,d) XRD linear scan patterns plotted in logarithm scale. Insets: optical images of the crystals with the crystallographic axes and dimensions indication. (e,f) XRD rocking curves taken on the [303] peak for the NbSiSb and [101] peak for the NbGeSb, respectively.

Similar to the ZrSiS, the NbGeSb and NbSiSb compounds possess a layered tetragonal crystal structure formed by stacking of Ge/Si-Nb-Sb-Nb-Ge/Si slabs along the *c* axis [Fig.

1(a,b)] [10, 11]. However, the shape of the as-grown single crystal is different from that of the ZrSiS or other *MXY* materials [inserts in Fig. 1(c,d)], which can be easily exfoliated to *c*-oriented thin flakes. The abnormal *c*-axis oriented needle shape can be attributed to the strong Nb-Si/Ge bonding. Shiny and smooth plane with the normal vector of [101], can be found from the single-crystal needles. XRD measurements were conducted on the shiny plane, resulting in sharp (*h0l*) patterns with *h*=*l*=1, 2, 3… [Fig. 1(c,d)]. The XRD rocking curves taken on the [303] diffraction peak of the NbSiSb as well as the [101] diffraction peak of the NbGeSb are shown in Fig. 1(e,f). The full width at half maximum (FWHM) of the rocking curve are ~0.18$^o$ and ~0.08$^o$, respectively, indicating good crystalline quality.

Taking the NbGeSb single crystal as an example, we employed SEM and HRTEM to demonstrate its microstructure. In Fig. 2, panel (a) demonstrates the EDS results taken on the shiny plane whose SEM image is shown in panel (b). Moreover, we employ the element mapping to reveal the distribution of the Nb, Ge, and Sb elements, which are plotted in panel (c-e). The obtained atomic ratios of Nb:Ge:Sb are 33.68 : 32.88 : 33.44. In order to further characterize the lattice structure and prove the high qualities of our crystals, we have carried out TEM measurement. We used the most conventional method of sample preparation, that is, grinding the crystals in the glove box, putting it into alcohol and ultrasonic treatment for 30 minutes, and then taking the upper liquid drop to the copper net for the final measurement. The results are shown in Fig. 2(f-k), the Energy-Dispersive X-ray mappings of NbGeSb via TEM give the similar results with that of SEM, both of them prove that our single crystal samples are of good quality and uniform distribution in both macroscopic and microscopic aspects. As pointed in Fig. 2(j), the HRTEM image of NbGeSb demonstrating nearly excellent crystalline and thus resulting in sharp SAED pattern [Fig. 2(k)] which can be indexed with the P4/nmm space group. All of these results evidence the excellent quality of the NbGeSb single crystals which provide a good platform for studying its electronic transport properties.

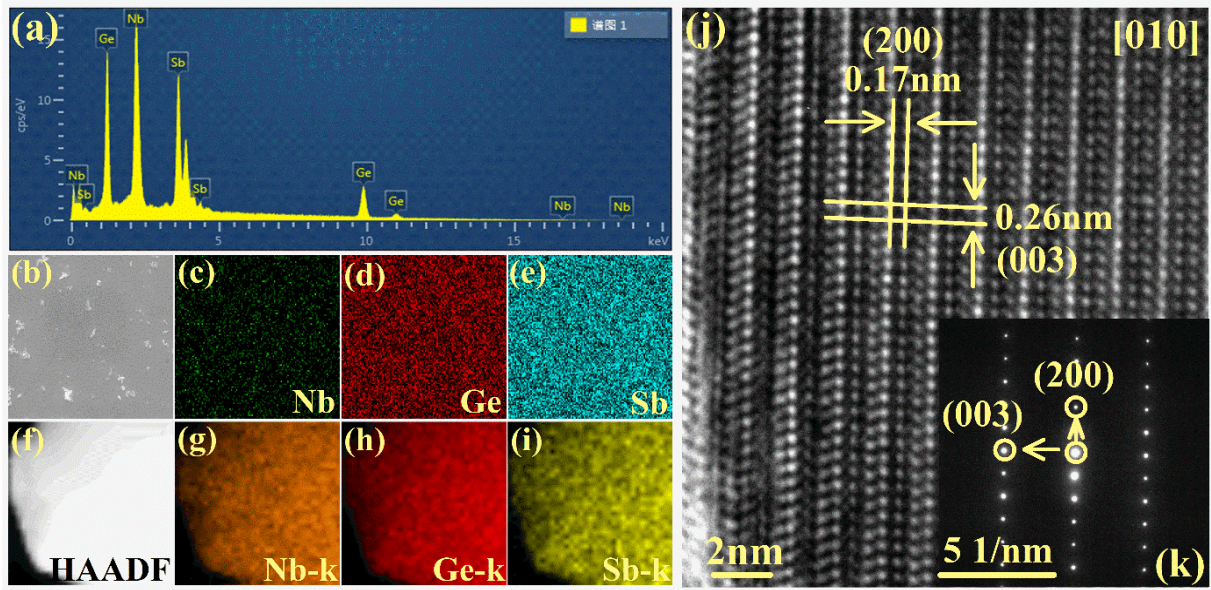

**Fig. 2** Electron microscopic properties of a NbGeSb single crystal. (a) the element index in energy dispersive x-ray spectroscopy. (b-e) SEM image (50×50 μm) and area-element (Nb, Ge, Sb) mapping of the NbGeSb's crystallographic *bc* plane. (f-i) Energy-dispersive X-ray mappings of NbGeSb via TEM. (j) HRTEM image of the NbGeSb lattice. (k) the selected area electron diffraction pattern with indices of NbGeSb.

## 4. Results and Discussion

Based on the crystal structure, first-principles calculations are performed first. The DFT optimized lattice constants are $a=b=3.67$ Å, $c=8.25$ Å for NbSiSb and $a=b=3.75$ Å, $c=8.30$ Å for NbGeSb, respectively. According to the density of states (DOS), the bulk band structure near the Fermi level are mainly contributed by the Nb's 4d orbitals for both materials, as shown in Fig. 3(a,b). The DOS near the Fermi level of NbGeSb is higher than that of NbSiSb, which implies metallic characteristics of NbGeSb is superior to that of NbSiSb. Furthermore, as shown in Figs. 3(e,f), the band structure confirms that multiple Fermi pockets (i.e. both electrons and holes) contribute to the transport properties of both NbSiSb and NbGeSb. In Figs. 3(c,d), we also plot the calculated 3D Fermi surface, here the blue part represents the electron pockets, and the yellow part represents the hole pockets, in which we can see some features: 1) the pockets on ΓY and ΓM direction are larger than ZrSiS family; 2) the linear-like dispersion bands can be found on ΓY and ΓZ directions; 3) comparing with ZrSiS, much more parabolic bands cross the Fermi level, introducing non-Dirac carriers into these crystals. One can see that the Fermi surfaces show strong anisotropy between *c* axis and *ab* plane, however within *ab* plane, four-fold symmetry with lower anisotropy (comparing with *bc* plane) can be found [8, 11]. In both compounds, Nb element contributes to the main density of states near Fermi level. Moreover, both compounds present open Fermi surface along the *a* direction, which may contribute to parabolic-like nonsaturating magnetoresistance and will be discussed later. We employ both the temperature- and angle-dependent electronic transport measurements to verify these band structure calculations.

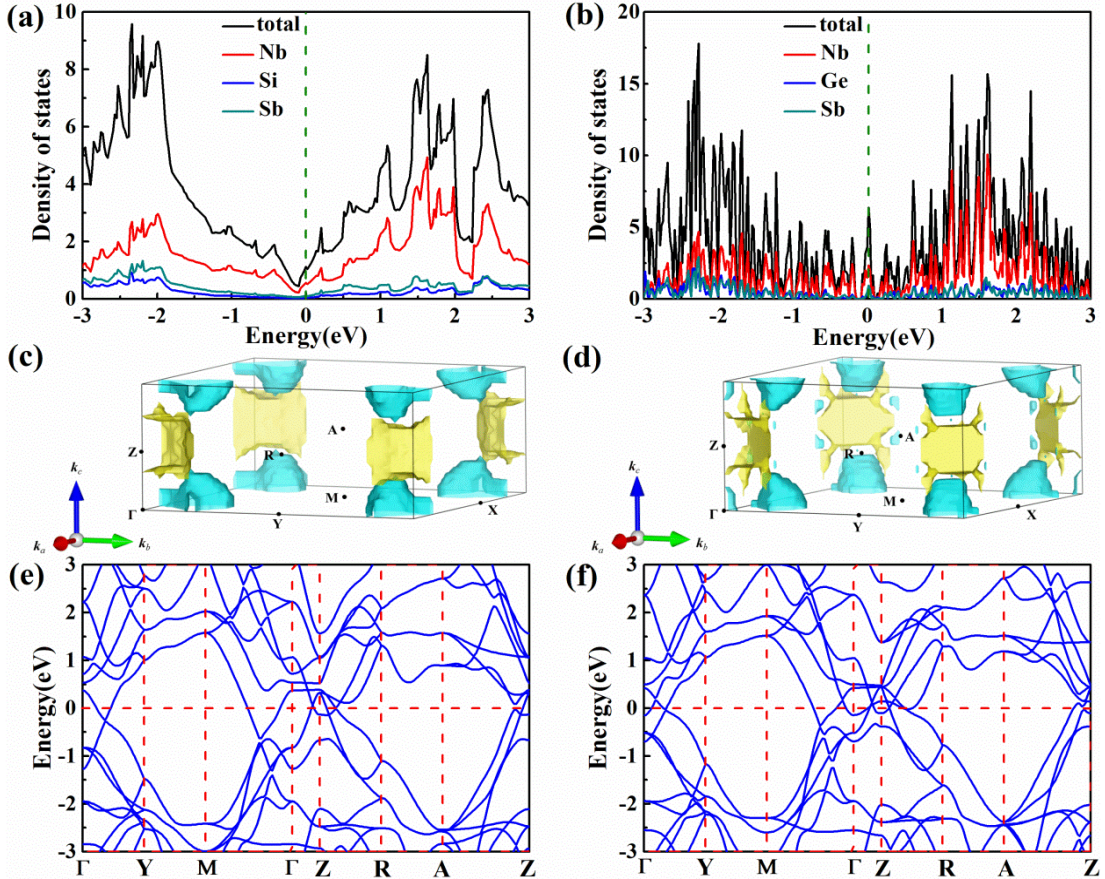

**Fig. 3** DFT calculation results of the electronic band structures. (a,b) density of states, (c, d) fermi surface, where the blue part represents the electron pockets, and the yellow part represents the hole pockets. (e, f) band structures. The inserts are 3D Fermi surfaces of the NbSiSb and NbGeSe in the reciprocal space, respectively.

As shown in Fig. 4 (a), the zero-field resistivity-temperature ($\rho$-$T$) curves of the NbSiSb and NbGeSb crystals show monotonically increasing within the entire temperature region (3-300 K), which suggests the metallic ground state of the NbSiSb and NbGeSb. Under a zero magnetic field the residual resistivity ratios (RRR) of the NbSiSb and NbGeSb are approximately 3.6 and 8.0, respectively, which is much lower than ZrSiS (~84) [11], slightly lower than ZrGeSe (14.9) [23], and similar to LaCuSb$_2$ (8.4) [24]. The entire $\rho T$ behavior of the NbSiSb and NbGeSb can be described in three different regions : 1) for temperature less than 10 K, the resistivity is almost constant for both single crystals, which are similar to the other semimetal systems [25, 26]; 2) between 10 K and 75 K, the resistivity can be quite well fitted using the equation $\rho(T)=\rho_0 +AT^n$ with $n$ ~2.4 for the NbSiSb, and ~2.49 for the NbGeSb, where $\rho_0$ is the residual resistivity, *A* is a constant, and *n* is a parameter indicating scattering mechanisms. This type of temperature dependence of resistivity, deviating from the pure electronic correlation-dominated scattering mechanism ($n$=2) observed in semimetal families, can be attributed to the interband electron-phonon scattering [23,27,28]; 3) for temperature higher than 80 K, the resistivity increases linearly with increasing temperature (the slopes and intercepts of linear fittings for NbSiSb are 1.8

μΩ·cm/K, -41.8 μΩ·cm; and for NbGeSb are ~3.3 μΩ·cm/K, 141.2 μΩ·cm respectively), which implies the pure electron-phonon scattering mechanism in this region (theoretically this linearly increasing would still work until structural phase transition) [29]. The slope of linear fitting is temperature coefficient of resistivity, which is the scattering intensity of degenerate electrons with phonons that are populated according to the classical equipartition distribution; and the intercept is the estimated residual resistivity.

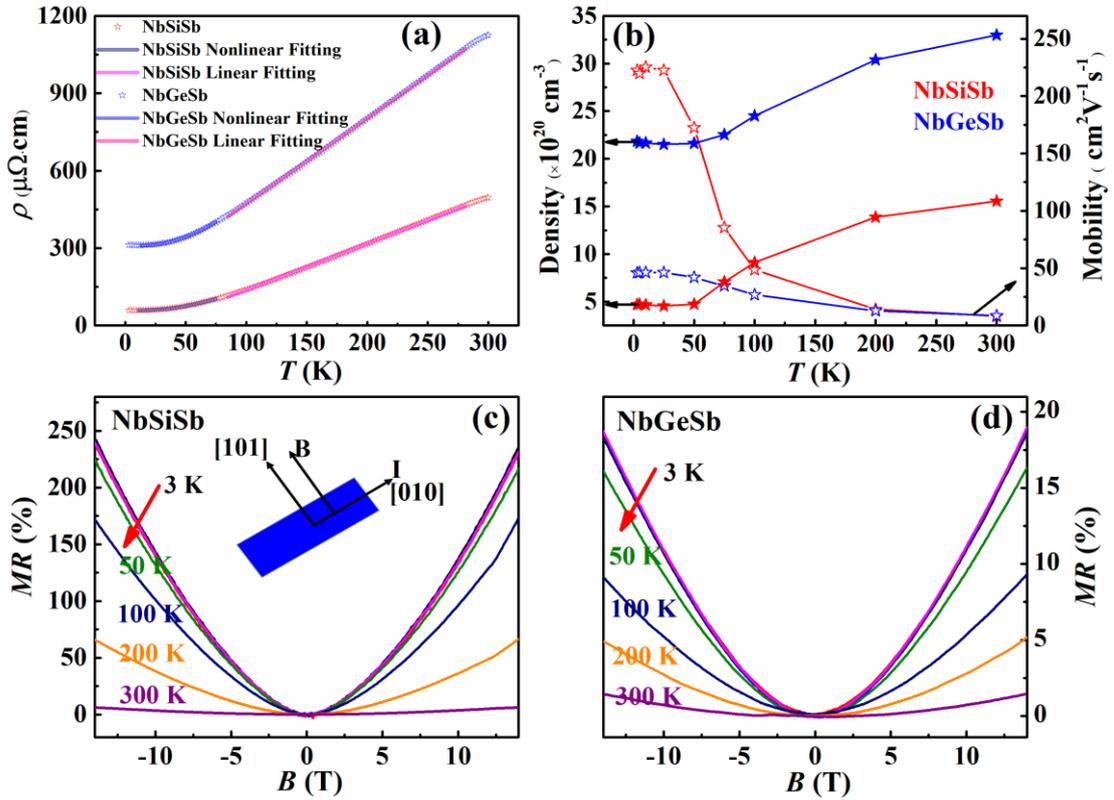

**Fig. 4** Electronic transport properties of the NbSiSb and NbGeSb single crystals. (a) resistivity as a function of temperature ($\rho T$). The low temperature $RT$ curve can be fitted using $\rho = \rho_0 + AT^n$. The high temperature $\rho T$ curve can be linearly fitted. (b) carrier density and mobility of NbSiSb and NbGeSb as a function of temperature. (c,d) the magnetoresistance of the NbSiSb and NbGeSb, respectively, where the direction of the magnetic field is along the [101] direction and the electric current is along the *b* axis.

We measured the temperature dependent transverse magnetoresistance (*MR*) which is the change of the resistance with the magnetic field applied perpendicular to the electric current. Here, $MR = \frac{\rho(B)-\rho(0)}{\rho(0)}$, where $\rho(B)$ is the resistivity in an applied magnetic field *B*. Since *MR* measurements could reflect the Fermi surface information of semimetals, we thus conducted the temperature dependent *MR* measurements of both samples at fixed temperatures ranging from 3 to 300 K and plotted *MR* as a function of magnetic field *B* in Fig. 4(c, d). For the NbSiSb, the low-temperature ($T \leq 50$ K) *MR* curves are parabolic like with large, non-saturating behavior, reaching ~240% at 3 K and 14 T. During heating, for $T \geq 100$ K the damping of *MR* value at 14 T becomes more significant because different scattering mechanisms (as aforementioned in *RT* discussion) work below and above $T \sim 75$ K. At 300 K, the *MR* is more-or-less neglectable since the electron-phonon scattering dominates the total transport behaviors. As aforementioned, the overall unsaturated MR behavior in NbSiSb is dominated by the open-orbit fermiology, agree with the DFT band calculations. The *MR* for the NbGeSb behaves almost the same as that of the NbSiSb, however with the maximum *MR* at 3 K and 14 T is one-order smaller (18%) than that of the NbSiSb. Due to the similar fermiology and lattice dynamic, the total scattering mechanism for the NbSiSb and NbGeSb are almost the same, resulting in similar *MR* behaviors [28, 29]. Since the MR behavior is proportion to the carriers' mobility, we deduce that the mobility difference in NbSiSb and NbGeSb finally resulting the different MR values. Therefore, we conduct the Hall effect measurements to show the carriers' properties in NbSiSb and NbGeSb, as shown in Fig. 3b. The total Hall effect shows linear-like decreasing curves, which correspond to a n-type conductor, are contributed by electron carriers (not included in the paper). From the Hall effect curves, we can calculate the Hall coefficient $R_H$, and thus obtained the carriers' density via $n_H = R_H^{-1} e^{-1}$ (*e* is the charge of an electron) and mobility $\mu_H = R_H \sigma$ ($\sigma$ is the conductivity). At 3 K, the carriers' density of NbSiSb and NbGeSb are $4.7 \times 10^{20}$ cm$^{-3}$ and $2.3 \times 10^{21}$ cm$^{-3}$, respectively. Together with the conductivity measured before, we can obtain the Hall mobility for NbSiSb and NbGeSb at 3 K are 230 and 48 cm$^2$/Vs, respectively. Comparing with ZrSiS, the carriers' density is one or two order's higher, and the mobility is also much lower, which probably contributed by the non-Dirac hole pockets on ΓZ, ZR directions.

To understand the geometry of the Fermi surface, we performed angle-resolved *MR* measurements. Note that, since the samples exhibit needle shape whose length direction is the *b* axis of the crystals. Therefore, the electric current direction flows along the *b* axis. It is easy to keep the direction of magnetic field perpendicular to the current direction during the rotation of the crystals within *ac* plane. However, it is hard to exclude the influences of magnetic field-current angle (θ) changing during the rotation from [101] direction to *b* axis. Here, it is assumed that the change in θ angle during the rotation of the crystals contributes much less than that contributed from fermiology. The schematic measurement diagrams are shown in the inset of corresponding angle-dependent *MR* panels (Fig. 5 a, b, c, d) . To clearly display the mirror symmetry (agree with the four-fold rotation symmetry within *ab* plane) between 0° - 90° rotation and 180° – 90° rotation, we manually added a negative sign on the MR values between 90° and 180°, as shown in Fig. 5. Let's focus on Panel (a), in which the rotation parameter *θ* is the angle between the [101] direction and the direction of the magnetic field, it can be seen that the angular *MR* shows less dependent on rotation within *ac* plane, which imply slightly in-plane (i.e., *ac* plane) anisotropy of the NbSiSb crystal. In contrast, when the direction of the magnetic field rotates from the [101] direction to *b*-axis direction, the Fermi surface changes significantly from the [101] direction to the *b* axis, resulting in a continuous change in MR upon rotation. For the NbGeSb crystal the rotation of the direction of magnetic field within *ac* plane results in a relatively larger in-plane *MR* anisotropy (Fig. 5c) with respect to that of the NbSiSb crystal. The rotation of the direction of magnetic field within *ac* plane has similar effects on the magnetotransport behaviors for the NbGeSb and NbSiSb [Fig. 5(b,d)]. In NbSiSb, the relatively higher mobility are mainly contributed by the linear-like bands on ΓM and ΓY directions, however in NbGeSb, the low mobility pockets on ΓZ and ZR direction. During the rotation measurements perpendicular to the current (*b* axis), in reciprocal lattice, the magnetic field rotates in Γ-Z-R plane, which probably results the relatively higher anisotropy in NbGeSb crystal than NbSiSb.

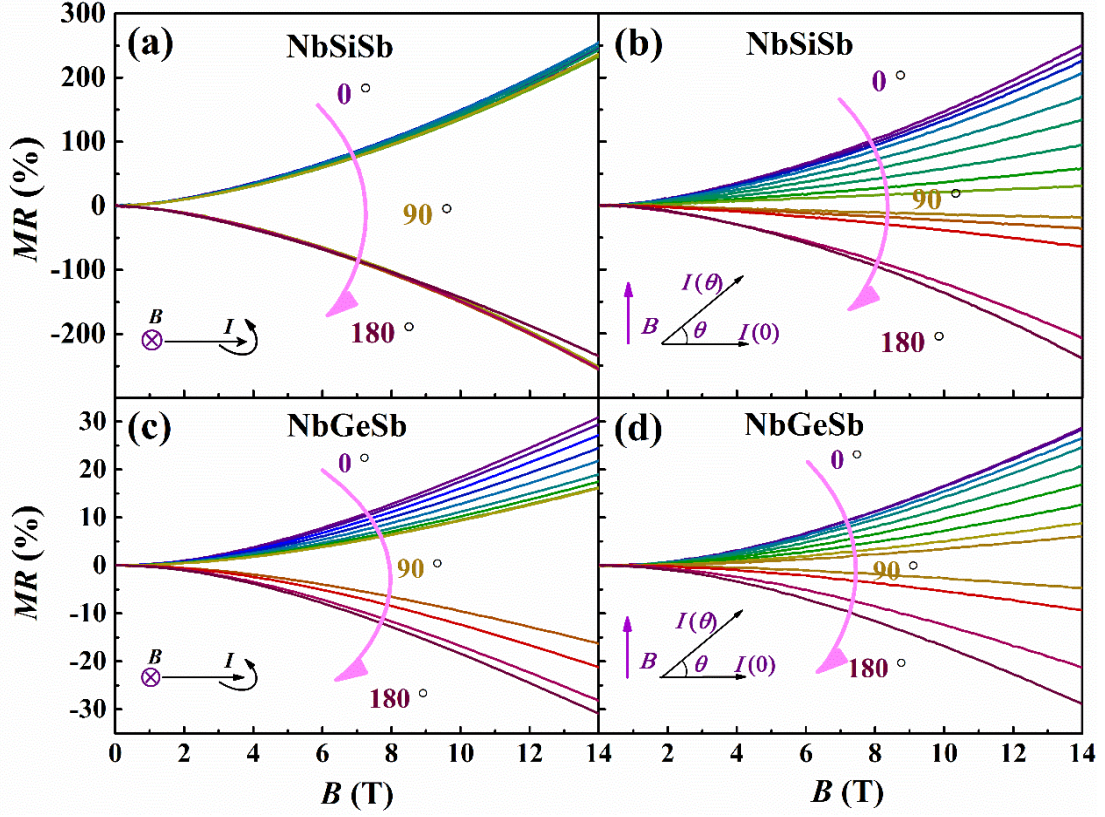

**Fig. 5** Angular dependent magnetotransport properties for the NbSiSb and NbGeSb crystals. (a, c) MR plot with the rotation angle, in which the initial direction of the magnetic field is along the [101] direction of the crystals while the electric current direction is along the *b* axis. During rotation the direction of the magnetic field was always kept perpendicular to the electric current direction. (b, d) MR plot with the rotation angle θ for the NbSiSb and NbGeSb crystals, respectively. Note that, in all diagrams, the *MR* values are positive, which means that the absolute values of *MR* are the real quantities. We plot the *MR* curve in two different regions to show the rotating effects.

We notice that in another semimetal system $PtSn_4$ and $PdSn_4$ [30]. Kohler's rule is successful to explain the field-induced low temperature resistance upturn upon cooling [31, 32]. Since Kohler's rule is both powerful and simplified, we employ Kohler's plot with temperature and angle dependent *MR* (which will be mainly discussed in next section) shown in Fig. 6. The Kohler's rule can be written with a scaling function *F*(x):

$$\Delta\rho/\rho_0 = F(H/\rho_0)$$

where $\Delta\rho = \rho(H) - \rho_0$, and $\rho_0$ is zero-field resistivity at a certain temperature. This relation follows from the fact that the magnetic field enters Boltzman's equation in the combination ($H\tau$) and that $\rho_0$ is proportional to the scattering rate $1/\tau$. Here, $\tau$ is the mean time between scattering events of the conduction electrons [17]. Such a scaling behavior expresses a self-similarity of the electronic orbital motion across different length scales: an invariance of the magneto-response under the combined transformation of the shrinking of the orbital length $L \sim 1/H$ by increasing $H$ while, at the same time also increase in the scattering rate $1/\tau$ by the same factor such that $H\tau$ remains unchanged, indicates that the system behaves identical on different length scales. Fig. 6 shows very similar behaviors between 3 and 300 K, and during rotation one more factor $\cos^{-1}(\theta)$ is considered to convert the magnetic field into effective magnetic field along the [101] direction, which may indicate similar magnetic scattering mechanism with the magnetic field at different temperature region [17].

Fig. 6 (a,b) demonstrate the temperature dependent Kohler's plot of the NbSiSb and NbGeSb, respectively, in which the *MR* curves are almost coincide with each other (the noncoincidence at low field region is due to the error amplification in log-log scale), especially in high temperature region. A similar magnetic-field-induced carriers' scattering mechanism can be used in the whole temperature region, regardless of the different zero-field scattering behavior observed in aforementioned $\rho T$ measurements. In angular dependent Kohler plots [Fig. 6 (c,d)], the same mechanism can be addressed if the effective magnetic field was converted to that along the [101] direction. Back to Fig. 3b, one can see the carriers' density and mobility stay nearly constant below 50 K, and changes quickly in high temperature region. However, the carriers' still behave the same in magnetic field, indicating that no phase transitions occur in NbSiSb or NbGeSb. The research of the electric transport behavior of NbSiSb and NbGeSb will be a good supplement to the family of nodal-line topological materials, and contribute to deep insight into topological behavior together with it's Fermi surface. Just as important, the highly similarity of crystallography and strong difference in band structure between the present single crystals and those of the ZrSiS family provides a possibility to tune the band structure via element doping.

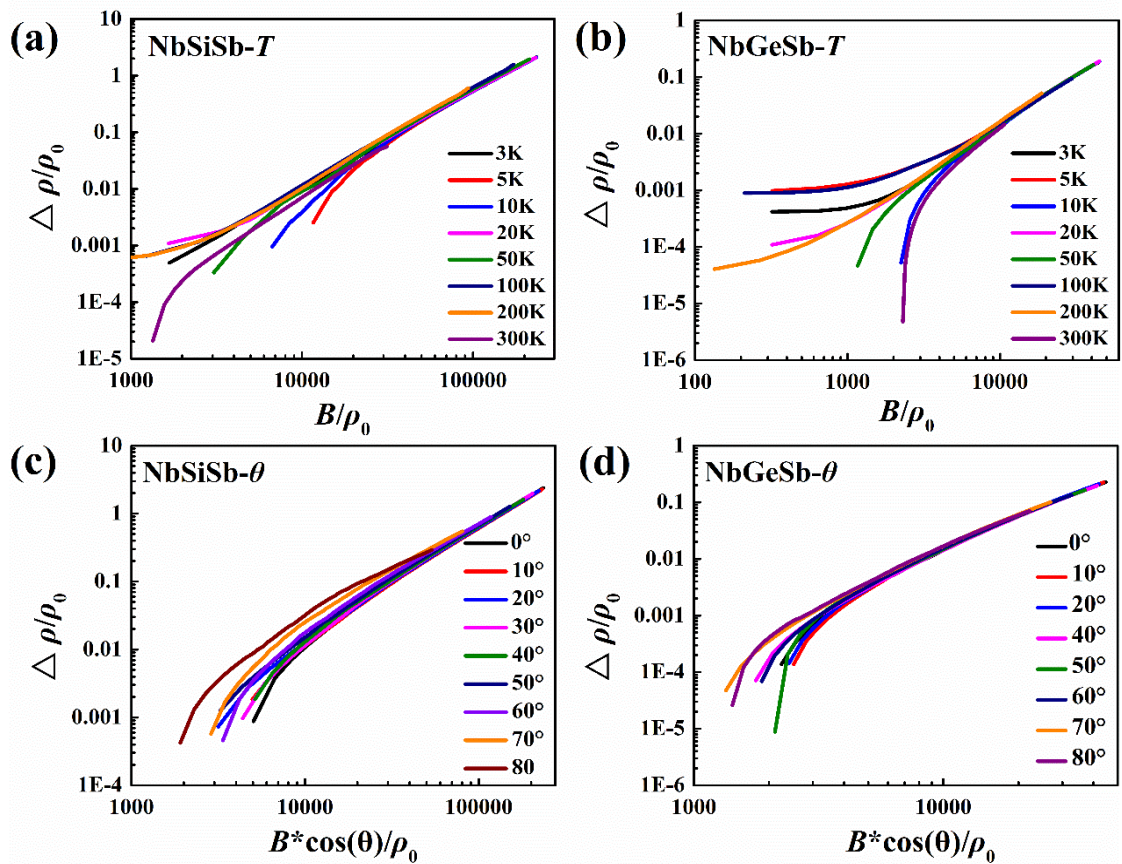

**Fig. 6** Kohler's plots on temperature and angle dependent magnetoresistance for the NbSiSb and NbGeSb crystals. (a, b) Kohler's plot with temperature dependent of the NbSiSb and NbGeSb. (c, d) Kohler's plot with angular dependent of the NbSiSb and NbGeSb, respectively.

## 5. Conclusions

In the present work, we employed the CVT method to grow needle-like single crystals of NbSiSb and NbGeSb which are isostructural materials with the nodal-line semimetals of the ZrSiS-family. The excellent crystalline quality is verified via single crystal XRD θ-2θ scan, rocking curve, SEM and HRTEM measurements. DFT calculations illustrate that in NbSiSb and NbGeSb, Nb elements contribute mainly to the DOS near Fermi level, including linear-like dispersed bands on ΓY, ΓM directions, as well as parabolic-like bands on ΓZ and ZR directions. In both crystals, the magnetoresistance show parabolic behavior with MR values of ~240% (NbSiSb) and ~18% (NbGeSb) at 3 K and 14 T. Hall measurements show that hole carriers dominant the transport behavior, with mobility at 3 K of 230 (NbSiSb) and 48 (NbGeSb) cm$^2$/Vs, respectively. The mobility difference in NbSiSb and NbGeSb directly results in different MR values, and probably affect the anisotropy during angular dependent MR measurements. Follows the Kohler's scaling rule, we prove that the similar inter- and intraband electron-phonon scattering mechanism plays a role in the NbSiSb and NbGeSb. Due to the bulk-dominant transport behavior in the present study, the evidence of NbGeSb's band topology is still absence. Topological transport behavior related to surface state Weyl orbital is still a challenge in NbGeSb, or related materials.

## Data Availability Statement

The data that support the findings of this study are available from the corresponding author upon reasonable request.

## Acknowledgements

This work was supported by the National Natural Science Foundation of China (Grant Nos. 11974155, 51572278, 51872278, 11674055). Support from Jiangxi Key Laboratory for Two-Dimensional Materials is also acknowledged. W.Z. & L.C. acknowledge the scholarship supporting from UOW.